\documentclass[journal,comsoc]{IEEEtran}
\usepackage{booktabs}
\usepackage{url}
\usepackage{graphicx}
\usepackage{cite}
\usepackage{xcolor}
\usepackage{tabularx}
\usepackage{pdflscape}
\usepackage{epsfig}
\usepackage{multirow}
\usepackage[ansinew]{inputenc}

\def\BibTeX{{\rm B\kern-.05em{\sc i\kern-.025em b}\kern-.08em
    T\kern-.1667em\lower.7ex\hbox{E}\kern-.125emX}}
    
\newcommand{\linebreakand}{%
  \end{@IEEEauthorhalign}
  \hfill\mbox{}\par
  \mbox{}\hfill\begin{@IEEEauthorhalign}
}

\begin{document}

\title{Digital Twin-enabled Intelligent DDoS Detection Mechanism for Autonomous Core Networks}

\author{Yagmur Yigit,~\IEEEmembership{Student Member,~IEEE,} Bahadir BAL,~\IEEEmembership{Member,~IEEE,}\\ Aytac Karameseoglu,~\IEEEmembership{Member,~IEEE,} Trung Q. Duong,~\IEEEmembership{Fellow,~IEEE},\\ and Berk Canberk,~\IEEEmembership{Senior Member,~IEEE}
\thanks{\textit{Y. YIGIT is within the Department of Computer Engineering, Faculty of Computer and Informatics, Istanbul Technical University; B. BAL is within the Microsoft-Denmark; A. KARAMESEOGLU is within the BTS Group, Istanbul-Turkey; T. Q. DUONG is a professor within School of Electronics, Electrical Engineering and Computer Science, Queen's University Belfast; B. CANBERK is a professor within the School of Engineering and Built Environment, Edinburgh Napier University, He is a professor within the Department of Artificial Intelligence and Data Engineering, Istanbul Technical University. Corresponding author is Trung Q. Duong.}}
}


\maketitle

\begin{abstract}
Existing distributed denial of service attack (DDoS) solutions cannot handle highly aggregated data rates; thus, they are unsuitable for Internet service provider (ISP) core networks. This paper proposes a digital twin-enabled intelligent DDoS detection mechanism using an online learning method for autonomous systems.  
Our contributions are three-fold: we first design a DDoS detection architecture based on the digital twin for ISP core networks. We implemented a Yet Another Next Generation (YANG) model and an automated feature selection (AutoFS) module to handle core network data.
We used an online learning approach to update the model instantly and efficiently, improve the learning model quickly, and ensure accurate predictions.
Finally, we reveal that our proposed solution successfully detects DDoS attacks and updates the feature selection method and learning model with a true classification rate of ninety-seven percent. Our proposed solution can estimate the attack within approximately fifteen minutes after the DDoS attack starts.
\end{abstract}


\section{Introduction}
Digital-twin (DT) concept has become a popular topic because of its benefits in many domains, such as real-time remote monitoring and control in industry, predictive maintenance in aerospace, etc. It is broadly envisioned that DT will play a significant role in developing ``zero-touch'' operations, maintenance, and ``self-X'' capabilities, e.g., self-management, self-configuration, self-optimization, etc., of 6G networks. Real-time network monitoring, performance testing, optimization, and fast simulation are just some examples that exploit the full advantages of DT in a network domain. Moreover, DT makes the evaluation, prediction, and optimization processes more cost-efficient than physical systems. Despite all the benefits of DT, the use of DT for network anomaly detection has not been well-understood \cite{dtsurvey}.

\begin{figure}[!t]
    \centering
    \includegraphics[width=3.3in]{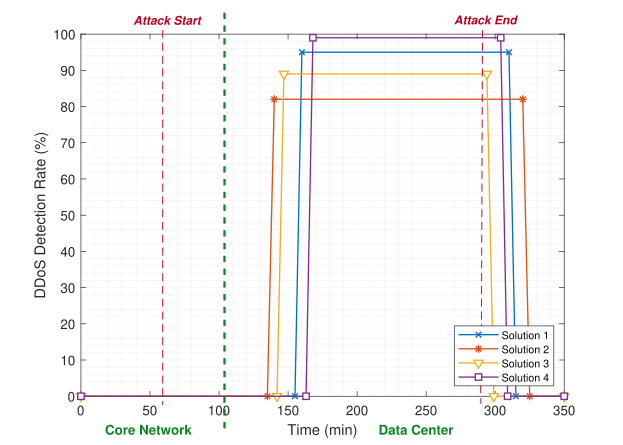}
    \caption{The detection performance of existing DDoS solutions.}
    \label{fig:bef}
\end{figure}

The main priority for an Internet service provider (ISP) is providing availability between content providers and end-users and lossless connectivity at high speed. ISPs manage massive traffic in their network, and distributed denial of service (DDoS) attacks are aimed at their critical network infrastructure and services. Existing DDoS solutions in the market are limited to data centers or edge networks. The solutions work on edge routers or inside the data center and do not protect other routers. For instance, even if the ISP activates the DDoS solution on 900 of 1000 edge network devices on a customer basis, the remaining 100 machines are still vulnerable. Therefore, data center and edge network solutions are insufficient; since it is impossible to incorporate DDoS detection in the entire network. 

We examined the DDoS solution of four different companies with our industry partner and anonymized them because of privacy as Solutions 1-4. Fig.~\ref{fig:bef} explains the solutions of detection rate and latency graphically; the left side of the green dashed line represents the core network, and the other side is the data center. When the DDoS attack starts at 60 minutes and ends at 290 minutes, these solutions detect approximately one hundred minutes after it begins. We found that the solutions' detection latency and rates are unsuitable for guaranteeing overall network performance.

\begin{table*}[!t]
\caption{ The summary of existing approaches
\label{tab:table1}} 
\centering
\begin{tabular}{l|cccc}
\toprule
\hline
\multicolumn{1}{c|}{\textbf{Work}}  & \textbf{\begin{tabular}[c]{@{}c@{}}Method\end{tabular}} & \textbf{Performance (\%)} & \textbf{Dataset} & \textbf{Domain} \\ \hline
\cite{tab6}  & Entropy-based Algorithm & 80 (Detection Rate) & Produced Dataset & SDN
\\
\cite{CIC2017}  & \begin{tabular}[c]{@{}c@{}} DeT, NB, RF, SVM, EM\end{tabular} & 74.1-99.4 (F1-Score) & CICDDoS2017 & IoT
\\
\cite{IF20}  &  Isolation Forest & 96.01 (F1-Score) & NSL-KDD & Fog Computing
\\
\cite{SmartGrid}  & \begin{tabular}[c]{@{}c@{}}Bagging, Boosting, Stacking\end{tabular} & 93/92.2/93.4 (Accuracy) & CICDDoS2019 & Smart Grid
\\
\cite{AEMLP}  &  MLP & 98.18 (F1-Score) & CICDDoS2019 & \begin{tabular}[c]{@{}c@{}} Intrusion Detection System \end{tabular}  
\\ 
\hline
\bottomrule
\end{tabular}
\end{table*}

\subsection{Main Challenges of The Core Network DDoS Detection}

We investigate the challenges from two main aspects in detail as follows:

\begin{enumerate}
    \item Specific Characteristics of The Core Network: The core network has high bandwidth because it includes routers with multiple 400 GBps interfaces. It makes it challenging to process the data in real-time. Therefore, ISPs are not interested in DDoS detection in the core network. To the best of our knowledge, there are no hardware-based solutions to solve it. Consequently, ISPs cannot control and manage DDoS attacks across their entire network. Some servers may have already been taken down when they detect DDoS attacks in the data center. ISP servers' average hourly downtime cost is between 300,000 and 400,000 U.S. dollars \cite{statista}. 
    Another challenge is core network routers handle much larger amounts of data simultaneously than other routers. Therefore, it only transmits data according to its routing tables and does not perform other data operations. 
    
    \item DDoS Attack: It is a security violation on a system where multiple distributed compromised machines target the victim server. It makes a server unavailable to legitimate users trying to access it. The distributed character of DDoS attacks makes them extremely difficult to counteract or find the source. It appears in various shapes and patterns, making it difficult to detect. Moreover, it can also perform easily by using the weaknesses of networks and by generating requests for services of the software. DDoS attacks are complicated to detect and mitigate in real-time, but DDoS detection is essential as these attacks can cause significant problems.
\end{enumerate}

\subsection{Why Do We Need DT, YANG Model and Online Learning for Autonomous Core Networks?}

Monitoring in the core network is quite complex as it involves an enormous number of metrics and multidimensional data to be tracked.
An internet service provider's primary responsibility is to ensure availability between content suppliers and end users, as well as lossless connectivity at high speeds. Internet service providers manage huge amounts of traffic in their networks, and attacks are targeted at vital network infrastructure and services.
Therefore, DT can be employed to handle monitoring and management of the core network with fully synchronous monitoring and management opportunity.
Moreover, DT is only implemented once, and online behavior analysis continuously updates itself. 
In contrast to traditional DDoS monitoring services, DT-enabled core networks can support services directly linked to network operational data rather than updated data.

Online learning is an ML method to create learning models using data in sequential order. It is used to predict future data at each time step. In classical ML approaches, called ``offline learning'', the whole data is used to train required models, and then the predictions are made on data points using the trained model.
Unlike offline learning, online learning updates the trained model. Thus, the model is entirely up to date with current events.
The main idea behind online learning is to update the model with new data when the performance of the trained model declines. 
In most ML-based DDoS detection models, the offline learning environment is used to train and test these models to detect DDoS attacks. The models may not achieve high accuracy when detecting actual DDoS attacks. 
Hence, existing ML-based DDoS detection models may not work well in real-time. Since the data in the core network is very diverse and high volume, offline learning models cannot handle this data with a stable performance. Therefore, online learning models can process the data with stable performance in the core network.

Network automation is a crucial topic for the entire networking industry in order to scale and move faster. Yet Another Next Generation (YANG) can be used to unlock the power of network automation \cite{yang2019}.
The YANG model is a network data model used to model configuration and state data manipulated by the Network Configuration Protocol (NETCONF).
It provides data modularity through modules and submodules and supports model-driven network automation.
The YANG model can reduce the complexity of the core network by pulling only the desired features. Thus, the system can work more effectively.

Considering the above advantages, we propose a DT-enabled DDoS detection mechanism using online learning and YANG model for autonomous core networks.
The YANG model will import only the desired features into the system.
DT will use a virtual representation of the ISP core network and learning capabilities to enable an intelligent DDoS detection mechanism.
Data captured from physical objects will represent high-level information integrating the behavior model of digitized objects in DT.

\begin{figure*}[!t]
    \centering
    \includegraphics[width=7.1in]{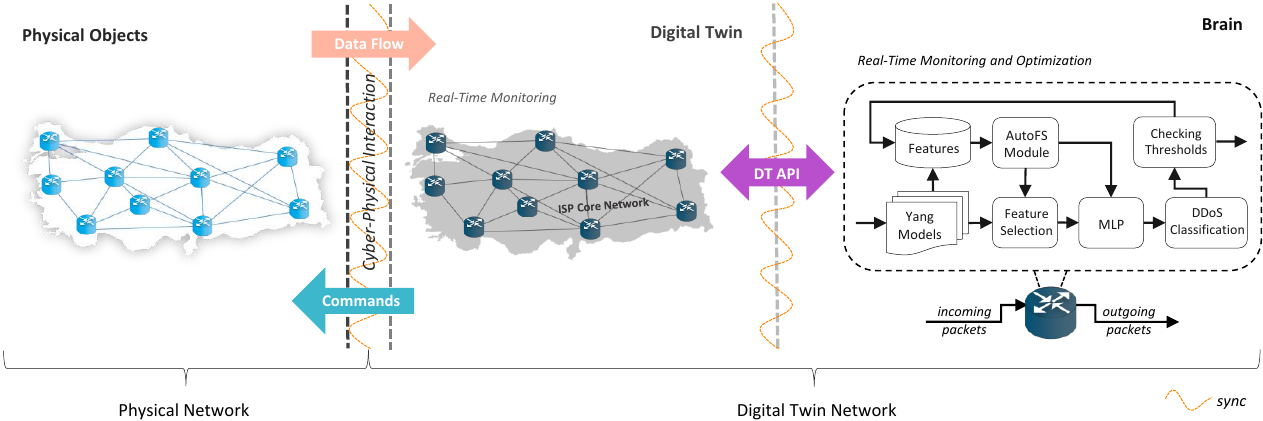}
    \caption{ The proposed system architecture.}
    \label{fig:system}
\end{figure*}

\subsection{Related Works}

Although there are many previous studies on DDoS detection, these studies generally focus on offline learning methods and data center solutions.
We summarized existing DDoS detection approaches in different domains in Table~\ref{tab:table1}. J. Boite \textit{et al.} proposed the StateSec is a DDoS detection and mitigation strategy based on stateful Software-Defined Networking (SDN) to protect communication endpoints \cite{tab6}. Simulation results showed that it is more efficient than sFlow for the control plane occupation. Z. K. Maseer \textit{et al.} compared several ML algorithms, including decision tree (DeT), naive Bayes (NB), random forest (RF), support vector machine (SVM), expectation-maximization (EM) in the perspective of DDoS detection \cite{CIC2017}. Experimental results showed that DeT and RF models achieve better than the others in terms of overall accuracy and runtime. In some studies, an autoencoder, a kind of neural network with multiple layers, is used for feature selection (FS).  Then, a classification algorithm is implemented to detect DDoS attacks \cite{IF20} \cite{AEMLP}. T. T. Khoei \textit{et al.} analyzed the performance of three different ensemble-learning techniques: bagging, boosting, and stacking for anomaly detection in smart grid networks. The results confirmed that stacking-based learning techniques outperformed the others.

So far, a few anomaly detection works are using DT in the literature. They focus more on offering the DT system rather than anomaly detection. For example, A. Saad \textit{et al.} suggested an Internet-of-Things (IoTs) based DT for the microgrids to enhance their resiliency against cyber-attacks \cite{dtref3}. They provided mathematical formulation and implementation of the DT. The results showed that their framework successfully mitigates the attacks.
Moreover, many works proposed a method to label unlabeled data. One of them offered a modified label propagation method \cite{Labeling}. The results showed that this method improved the fault classification accuracy effectively.
None of the above works focused on DDoS detection using online learning and the YANG model, nor did they use DT technology to detect DDoS attacks in the ISP core network.

\subsection{Contributions}

The choice of the classification algorithm is significant for DDoS detection. We scrutinized many ML algorithms considering the training and testing times.
We found that using multilayer perceptron (MLP) in the proposed solution performs better for the system.
This paper proposes a DT-enabled DDoS detection mechanism using online learning for autonomous core networks. Moreover, we suggest an automated FS (AutoFS) module to reduce and find the most appropriate features. Our proposed system architecture is shown in Fig.~\ref{fig:system}, where the physical objects, the digital twin, and the brain parts work in synchronization and communicate in real-time. 
Furthermore, the proposed architecture is in line with the Internet Engineering Task Force (IETF) digital twin network concept \cite{ietf}.

The main contributions of this paper can be summarized as follows:
\begin{itemize}
    \item We propose an intelligent DDoS detection mechanism taking advantages of DT for autonomous core networks.
    \item We implement an online learning technique to maintain stable performance since data are taken sequentially and make the best predictions for future data at each step. Moreover, we propose a labeling algorithm to label unlabeled data.
    \item We implement the YANG model and AutoFS module to reduce complexity and find the most relevant features in the core network. Thanks to them, our system can work separately for each router.
\end{itemize}

The rest of the paper is organized as follows. The proposed solution and the performance evaluation are explained in Section~\ref{sec:proposed} and Section~\ref{sec:perfEval}, respectively. We conclude the paper in Section~\ref{sec:conc} with future directions.

\section{Proposed Solution}
\label{sec:proposed}

The proposed mechanism detects DDoS attacks on a router-by-router basis, and it works as follows:

\begin{itemize}
\item A replica of physical objects and synchronization between the physical objects, the DT, and the brain are created.
\item DT collects all data from the core network routers, which are physical objects.
\item Data of digitized objects in DT are imported into the system through a YANG model. 
\item Thanks to the YANG model, the ninety-two features are sent to the system FS method.
\item The ten best features are sent to the system MLP as input.
\item The system MLP decides if there is a DDoS attack on the system. 
\item After that, the threshold values of the performance metrics are checked. 
\item If one of the performance metrics is less than its threshold values, the features used by the AutoFS module are updated. 
\item In the AutoFS module, one thousand samples are selected randomly for the five FS methods separately in current data. Each FS method determines the ten best features according to its algorithm.
\item The proposed labeling algorithm labels the data.
\item Then, the labeled data is used in MLP for training and testing.
\item The final FS algorithm chooses the best FS method and MLP model for the data.
\item After that, the system FS method and MLP model are updated according to the AutoFS module result.
\item If the performance metrics are not lower than their threshold, the system operates with the current FS method and MLP model.
\end{itemize}

The flow of the proposed detection mechanism is shown in the brain part in Fig.~\ref{fig:system}. All decisions are made in the brain part of the system. The performance of our approach is evaluated with four metrics: detection performance, system precision, sensitivity, and F-measure.  

\subsection{YANG Model}

YANG models comprise modules and submodules. It can define configuration and state data, notifications, and Remote Procedure Calls (RPCs) for use by NETCONF-based operations.
Some key YANG model capabilities are human readable, easy to learn representation, hierarchical configuration data models, reusable types and groupings (structured types), data modularity through modules and submodules, and extensibility through augmentation mechanisms \cite{rfc7950}.
There are many benefits to employing the YANG model in the network. Some of them are faster service deployment enabled by automation, improved serviceability, faster time to diagnose and repair issues, reduced operating costs by reducing legacy network engineering expenses, etc. \cite{yang2019}.
It enables the transition from command-line-interface-based management toward data model-driven management. Since the sensor data in the core network routers is more than the other routers, receiving all data makes the system very slow. It is necessary to reduce data being imported into the system. Therefore, we use a YANG model to accomplish this. 

We define two key performance indicators (KPIs) for DDoS attack detection with our industry partner. These KPIs are KPI-1 and KPI-2. KPI-1 contains 37 sensors, and KPI-2 has 55 sensors. We use a YANG model to get the data from the KPIs into our system.
Thanks to YANG Paths, the proposed system receives data from all 92 sensors from routers in the core network. 
Thus, we get less data into the system and increase the system's efficiency.  
The hierarchy between terms is as follows: sensor data received from routers via YANG Paths creates YANG models, and YANG models comprise KPIs.

\begin{figure*}[!t]
    \centering
    \includegraphics[width=7.1in]{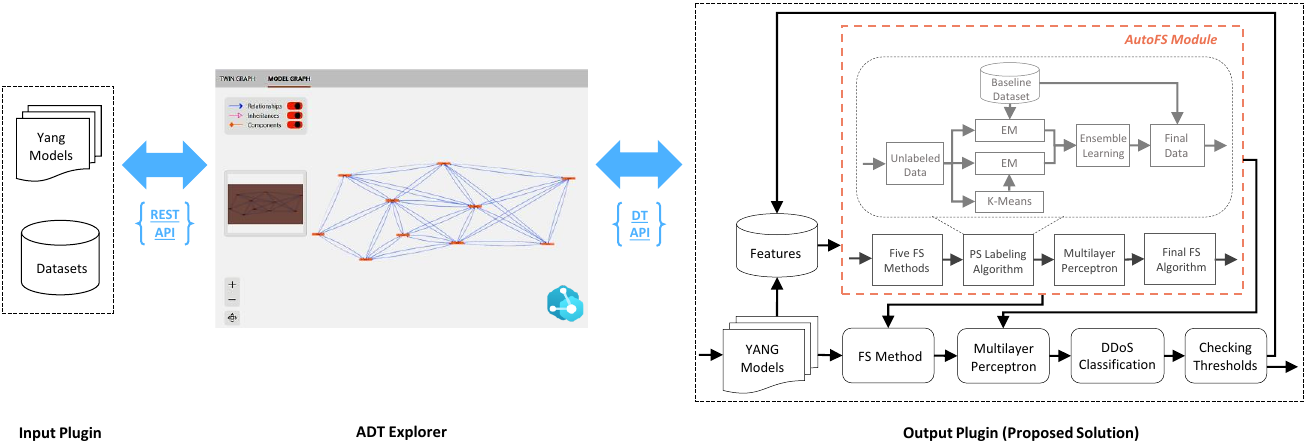}
    \caption{The visualization of the proposed solution through the Microsoft Azure Digital Twins platform.}
    \label{fig:twin}
\end{figure*}

\subsection{AutoFS Module}
We examined many FS methods and decided on the five FS methods that are most suitable for our system. AutoFS module determines the best FS method for the system among Analysis of Variance (ANOVA) F-value Selection, Chi-square, Backward Feature Elimination (BFE), Fisher Score, and Recursive Feature Elimination (RFE).
Moreover, since network data is not truly stable and available online, we used different FS techniques to give different results for diverse data.
ANOVA F-value Selection is a statistical hypothesis test that detects the feature impact on the class variable. It uses the f-tests to test the equality of means statistically.
Chi-square is a statistical method in discrete data testing that evaluates two variables, whether these variables are related or not. It selects the features discarding irrelevant features. 
BFE starts the search with the complete set of features and eliminates them one by one using some form of feature scoring until an optimal subset is found.
Fisher Score selects each feature independently according to their scores and then ranks by their importance.
RFE performs a greedy search to find the best-performing feature subset. It ranks features and recursively eliminates a small number of elements per loop.
When we examined these methods, we observed that they behave differently according to network traffic. While ANOVA F-value Selection and RFE are most suitable for real-time traffic, BFE and Fisher Score are ideal for non-real-time traffic. Chi-square fits for best-effort traffic.

When one of the system's performance metrics is less than its threshold values, the AutoFS module decides the best FS method and MLP model for the system. It takes a thousand random samples from the current data. This data is then used as input by the five FS methods.
After reviewing the number of parameters used for DDoS detection in the literature, we defined the optimum number of features as ten. The FS techniques select the ten best features according to their algorithm. Since existing data is unlabeled, we need to label it.
Therefore, we propose a labeling algorithm to label existing data. 

\subsubsection{Proposed Labeling Algorithm}
We used an ensemble learning algorithm that combines K-Means and EM algorithms. The flow of the proposed labeling algorithm is shown in the output plugin part in Fig.~\ref{fig:twin}. The proposed labeling algorithm works as follows:

\begin{itemize}
    \item The unlabeled data is taken as input.
    \item Firstly, we defined K as equal to two for the K-Means algorithm. It clusters the data into two groups and is used to determine the range of EM initial values. Thus, we improved the convergence speed and the stability of the EM algorithm.
    \item Then, the EM is applied to assign the weighted labels probabilistic for unlabeled data.
    \item Since DDoS attacks rarely occur, we created a baseline dataset to predict labels correctly. After our calculations, we defined the samples of the baseline dataset as one thousand. It includes sixty-five percent of `DDoS Attack' samples.
    \item The other EM algorithm uses both the baseline data and the unlabeled data. It finds local maximum likelihood estimation of parameters in its probabilistic models.
    \item Ensemble learning takes both the first and second EM algorithm results as input. Then, it decides the final labels of the data.
    \item Finally, the data, which is the output of the ensemble learning, and the baseline data are federated.
\end{itemize}

After that, we have the labeled data with two thousand samples and ten features. The labeled data is used for the training and testing of MLPs. Then, the five techniques' recall and detection time values are sent to the final FS algorithm. This algorithm decides the best FS method by optimizing each method's recall and detection time. It finds the best approach with maximum recall and optimal detection time for the system. Then, the system FS method and MLP model weights are updated according to the best FS method.

\subsection{Multilayer Perceptron}
MLP is a feedforward neural network consisting of several hidden layers where each hidden layer's output becomes the input of the consecutive layers. We investigated several MLP architectures. The results showed that the best optimized MLP has five layers: one input layer, three hidden layers, and one output layer. We also examined different activation functions. The best activation function was Rectified Linear Unit (ReLU) for hidden layers and softmax for the output layer. 
Therefore, our MLP has these parameters.
In addition, we used the dropout method to reduce the overfitting of MLP. 

The network data constantly flows and varies; it is not fixed as in offline learning. Therefore, we used online learning for the core network. We provide it by checking the threshold of performance metrics. 
If one of them is less than the threshold values, it indicates that our MLP model performance is low, and we need to update the model's weights. 
Thanks to the AutoFS module, we update the weights of our system MLP to ensure stable performance. 
Thus, we provide online learning updating the weights of the MLP according to the network status. 
MLP classifies data as `DDoS' and `Not DDoS'.

\subsection{The Complexity of the System}
Since we used the YANG model and the AutoFS module, we first reduced the number of features to ninety-two and then ten. Thus, we have significantly reduced the system complexity. Since we predefined all metrics, including the number of samples, features, layers, neurons, etc., in the AutoFS module, the complexity of AutoFS can be regarded as a constant. Therefore, the worst classification runtime has an upper bound that depends on the packet arrival rate and the number of core network routers.

\section{Performance Evaluation}
\label{sec:perfEval}

We use Microsoft Azure DT (ADT), a platform as a service tool to provide twin graphs of the physical objects \cite{Azure}. The capabilities of ADT are DT Definition Language (DTDL), an open modeling language, live representation, input plugin, and output plugin. 
We defined the digital models representing our physical entities, core network routers, using DTDL. These models identify semantic relationships between the entities. Therefore, we connected the twins into a graph that reflects their interactions. We designed a proof of concept model to test our system, and the visualization of the twin graph can be seen in Fig.~\ref{fig:twin} through ADT. The data in the ADT platform can stream through an external output plugin for storage and data analysis. After implementing the twin model and graphs with predefined interfaces and relations among entities, we transmitted data to the brain as the output plugin of ADT using the DT API. We designed the brain part of the proposed solution (PS) as microservice-based.

We used two different datasets to test PS as the input plugin of ADT. The first dataset is the CICDDoS2019 dataset, which resembles real-world data containing benign and the most recent common DDoS attacks \cite{dataset1}. This dataset includes eighty-five features and will be named D1 hereafter. The second dataset is the ToN IoT dataset created to evaluate the fidelity and efficiency of AI-based cybersecurity applications for next-generation IoTs and industrial IoTs \cite{dataset2}. It contains 43 features and will be named D2 from now on.
After examining the datasets, we found that they were imbalanced. D1 has 5159863 samples as DDoS attacks and 1502 samples as Not DDoS. Similarly, D2 has 300000 samples as DDoS attacks and 161043 samples as Not DDoS. Therefore, to nearly balance D1, we firstly used random undersampling by twenty percent and decreased the Not DDoS samples. After that, we used the synthetic minority oversampling technique (SMOTE). To almost balance D2, we applied the near-miss undersampling method and decreased the Not DDoS samples. Thus, we obtained more balanced datasets. We combined these two datasets to test PS. We used stratified ten-fold cross-validation to split the dataset randomly, maintaining the same class distribution in each subset. We also prepared a baseline dataset to test the AutoFS module. 

\begin{figure}[htbp]
    \centering
    \includegraphics[width=3.5in]{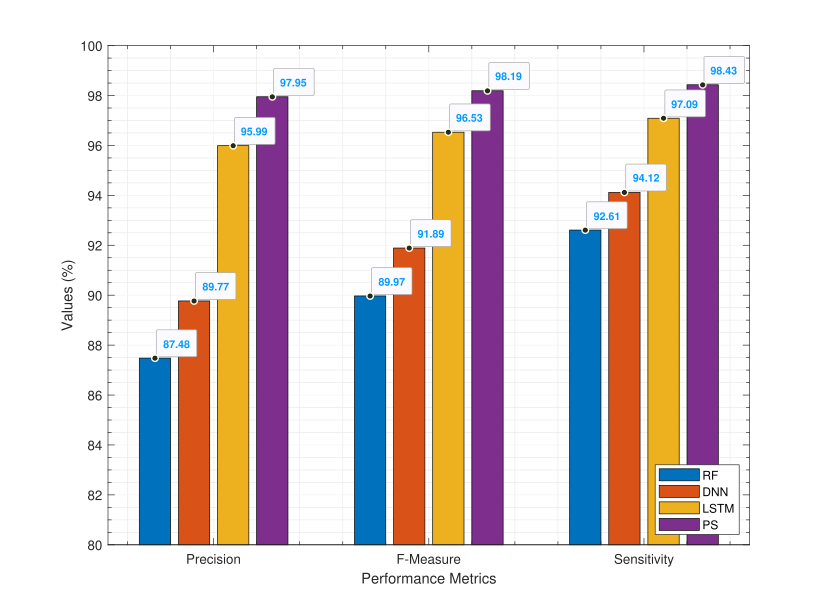}
    \caption{Performance comparison of methods.}
    \label{fig:pm}
\end{figure}

We compared PS with several algorithms: RF, a simple deep neural network (DNN), and long short-term memory (LSTM), a deep learning architecture.
Simple DNN has five hidden layers and the number of neurons per layer: 8191, 4096, 2048, 1024, and 512. 
LSTM has four fully connected layers and a number of neurons per layer: 256, 128, 64, and 32. 
We calculated the system precision, F-measure, and sensitivity values by taking the weighted average.
We used sixty percent of the DDoS class and forty percent of the Not DDoS class. Fig.~\ref{fig:pm} depicts the performance result of the methods.
Since PS uses online learning and can update the model, it performs better than the others. 
After that, we examined the performance of the AutoFS module and the online learning method.
The results prove that PS successfully updates the FS method and the weights of the MLP with a 97 \% true classification rate. 

\begin{figure}[htbp]
    \centering
    \includegraphics[width=3.3in]{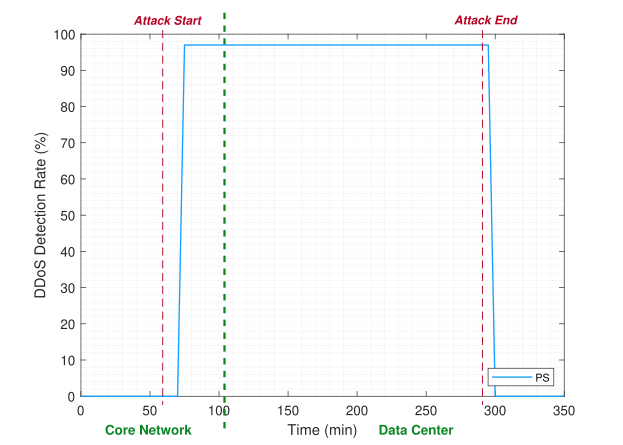}
    \caption{The detection performance of Proposed Solution.}
    \label{fig:after}
\end{figure}

According to our calculations, PS can estimate the attack approximately fifteen minutes after the DDoS attack starts. Moreover, its detection rate is around ninety-seven percent. Fig.~\ref{fig:after} exposes the detection performance of PS. When Fig.~\ref{fig:bef} and Fig.~\ref{fig:after} are compared, it can clearly see that PS outperforms the existing solutions regarding the detection delay and rate.

\section{Conclusion \& Future Directions}
\label{sec:conc}
In conclusion, we designed an intelligent DDoS detection mechanism based on DT technology for autonomous core networks.  
The proposed system detects DDoS attacks using only the required data obtained from the YANG model and AutoFS module, ensuring the computational capacity of the model can be sustained by the network. 
Finally, we showed that our solution successfully detects DDoS attacks using the ideal feature selection method for the network. It also updates the feature selection method and learning model with a ninety-seven percent true classification rate. Moreover, according to our analysis, the proposed solution can estimate the attack within around fifteen minutes after the DDoS attack starts.

As one of the future research directions, the router health of the core network can be scrutinized and designed as a monitoring module. This module can add proposed architecture to provide comprehensive core network management. To this end, Interior Gateway Protocols (IGP), especially Intermediate System to Intermediate System (IS-IS) and Open Shortest Path First  (OSPF), can be investigated for router health.

\section*{Acknowledgment}
Yagmur Yigit would like to thank the DeepMind Scholarship Programme and the ITU-Turkcell Graduate Research Scholarship Programme for their support. This paper is supported by Istanbul Technical University, Department of Scientific Research Projects with project number ITU-BAP, MCAP-2022-43826.

\bibliographystyle{IEEEtran}

\begin{thebibliography}{10}
\providecommand{\url}[1]{#1}
\csname url@samestyle\endcsname
\providecommand{\newblock}{\relax}
\providecommand{\bibinfo}[2]{#2}
\providecommand{\BIBentrySTDinterwordspacing}{\spaceskip=0pt\relax}
\providecommand{\BIBentryALTinterwordstretchfactor}{4}
\providecommand{\BIBentryALTinterwordspacing}{\spaceskip=\fontdimen2\font plus
\BIBentryALTinterwordstretchfactor\fontdimen3\font minus
  \fontdimen4\font\relax}
\providecommand{\BIBforeignlanguage}[2]{{%
\expandafter\ifx\csname l@#1\endcsname\relax
\typeout{** WARNING: IEEEtran.bst: No hyphenation pattern has been}%
\typeout{** loaded for the language `#1'. Using the pattern for}%
\typeout{** the default language instead.}%
\else
\language=\csname l@#1\endcsname
\fi
#2}}
\providecommand{\BIBdecl}{\relax}
\BIBdecl

\bibitem{dtsurvey}
Y.~Wu, K.~Zhang, and Y.~Zhang, ``{Digital Twin Networks: A Survey},''
  \emph{IEEE Internet of Things Journal}, vol.~8, no.~18, pp. 13\,789--13\,804,
  May 2021.

\bibitem{tab6}
J.~Boite, P.-A. Nardin, F.~Rebecchi, M.~Bouet, and V.~Conan, ``{Statesec:
  Stateful monitoring for DDoS protection in software defined networks},'' in
  \emph{IEEE Conference on Network Softwarization (NetSoft)}, Bologna, Italy,
  July 2017, pp. 1--9.

\bibitem{CIC2017}
Z.~K. Maseer, R.~Yusof, N.~Bahaman, S.~A. Mostafa, and C.~F.~M. Foozy,
  ``{Benchmarking of Machine Learning for Anomaly Based Intrusion Detection
  Systems in the CICIDS2017 Dataset},'' \emph{IEEE Access}, vol.~9, pp.
  22\,351--22\,370, Feb. 2021.

\bibitem{IF20}
K.~Sadaf and J.~Sultana, ``Intrusion detection based on autoencoder and
  isolation forest in fog computing,'' \emph{IEEE Access}, vol.~8, pp.
  167\,059--167\,068, Sept. 2020.

\bibitem{SmartGrid}
T.~T. Khoei, G.~Aissou, W.~C. Hu, and N.~Kaabouch, ``{Ensemble Learning Methods
  for Anomaly Intrusion Detection System in Smart Grid},'' in \emph{2021 IEEE
  International Conference on Electro Information Technology (EIT)}, July 2021,
  pp. 129--135.

\bibitem{AEMLP}
Y.~Wei, J.~Jang-Jaccard, F.~Sabrina, A.~Singh, W.~Xu, and S.~Camtepe, ``Ae-mlp:
  A hybrid deep learning approach for ddos detection and classification,''
  \emph{IEEE Access}, vol.~9, pp. 146\,810--146\,821, Oct. 2021.

\bibitem{statista}
T.~Alsop. {Average cost per hour of enterprise server downtime worldwide in
  2019}. {[Online]. Available}:
  \url{https://www.statista.com/statistics/753938}, Accessed Sept. 9, 2021.

\bibitem{yang2019}
B.~Claise, J.~Clarke, and J.~Lindblad, \emph{{Network Programmability with
  YANG: The Structure of Network Automation with YANG, NETCONF, RESTCONF, and
  gNMI}}.\hskip 1em plus 0.5em minus 0.4em\relax Addison-Wesley Professional,
  2019.

\bibitem{dtref3}
A.~Saad, S.~Faddel, T.~Youssef, and O.~A. Mohammed, ``{On the Implementation of
  IoT-Based Digital Twin for Networked Microgrids Resiliency Against Cyber
  Attacks},'' \emph{IEEE Transactions on Smart Grid}, vol.~11, no.~6, pp.
  5138--5150, June 2020.

\bibitem{Labeling}
Y.~Xie, ``{Modified Label Propagation on Manifold With Applications to Fault
  Classification},'' \emph{IEEE Access}, vol.~8, pp. 97\,771--97\,782, May.
  2020.

\bibitem{ietf}
C.~Zhou, H.~Yang, X.~Duan, D.~Lopez, A.~Pastor, Q.~Wu, M.~Boucadair, and
  C.~Jacquenet. {Digital Twin Network: Concepts and Reference Architecture}.
  {[Online]. Available}:
  \url{https://datatracker.ietf.org/doc/html/draft-zhou-nmrg-digitaltwin-network-concepts-06},
  Accessed Dec. 15, 2021.

\bibitem{rfc7950}
M.~Bjorklund. {The YANG 1.1 Data Modeling Language, RFC 7950}. {[Online].
  Available}: \url{https://www.rfc-editor.org/rfc/rfc7950.html}, Accessed Aug.
  12, 2022.

\bibitem{Azure}
Microsoft. {Azure Digital Twins Documentation}. {[Online]. Available}:
  \url{https://docs.microsoft.com/en-us/azure/digital-twins}, Accessed Sept.
  27, 2021.

\bibitem{dataset1}
I.~Sharafaldin, A.~H. Lashkari, S.~Hakak, and A.~A. Ghorbani. {DDoS Evaluation
  Dataset (CIC-DDoS2019)}. {[Online]. Available}:
  \url{https://www.unb.ca/cic/datasets/ddos-2019.html}, Accessed June 21, 2021.

\bibitem{dataset2}
N.~Moustafa. {ToN IoT datasets}. {[Online]. Available}:
  \url{https://ieee-dataport.org/documents/toniot-datasets}, Accessed June 21,
  2021.

\end{thebibliography}


\newpage

\section{Biography Section}

\vspace{-105pt}

\begin{IEEEbiographynophoto}{Yagmur Yigit}
[S'21](yigity20@itu.edu.tr) is an MSc student at the Department of Computer Engineering, Faculty of Computer and Informatics, Istanbul Technical University (ITU), Istanbul, Turkey. She is also a DeepMind scholar and part of a broadband communication and network automation research group under ITU. Her research focuses on AI-assisted network management.
\end{IEEEbiographynophoto}

\vspace{-105pt}
\begin{IEEEbiographynophoto}{Bahadir Bal}
(Bahadir.Bal@microsoft.com) received his BSc in Computer Engineering from ITU in 2014. He is currently working as a software engineer in the Microsoft development center of Copenhagen. He is also part of the broadband communication and network automation research group under ITU. His current research areas include 6G analysis, digital-twin networks, and their relations with Industry 4.0.
\end{IEEEbiographynophoto}

\vspace{-105pt}

\begin{IEEEbiographynophoto}{Aytac Karameseoglu}
(akarameseoglu@btsgrup.com) received a BSc in Chemical Engineering at Istanbul University but decided to switch his technical knowledge to learn Computer and Network Engineering. He has over 20 years of experience in the Networking Industry based on Cisco System Solutions and Devices. He achieved various Cisco Certifications and CCIE Certification (CCIE18102) in 2007, Cisco Champion in 2019. His current research areas include Segment Routing, Segment Routing v6 Programmability, Automation and Orchestration in Massively Scalable Infrastructures.
\end{IEEEbiographynophoto}

\vspace{-105pt}

\begin{IEEEbiographynophoto}{Trung Q. Duong}
[S'05, M'12, SM'13, F'22] (trung.q.duong@qub.ac.uk)  currently a chair professor of telecommunications at Queen's University Belfast, U.K., and a Research Chair of the Royal Academy of Engineering. His current research interests include wireless communications, signal processing, machine learning, and data analytic with the applications in environment, disaster management, healthcare, agriculture, and industrial automation. He was awarded the Best Paper Award at the IEEE Vehicular Technology Conference (VTC-Spring) in 2013, the IEEE International Conference on Communications (ICC) 2014, the IEEE Global Communications Conference (GLOBECOM) 2016, the IEEE Digital Signal Processing Conference (DSP) 2017, and GLOBECOM 2019. He was a recipient of prestigious Royal Academy of Engineering Research Fellowship from 2015 to 2020 and has won a prestigious Newton Prize 2017. He also serves as an Editor for the Scientific Reports (Nature), the IEEE Transactions on Wireless Communications, and the IEEE Transactions on Vehicular Technology; and an Executive Editor for IEEE Communications Letters.
\end{IEEEbiographynophoto}

\vspace{-105pt}

\begin{IEEEbiographynophoto}{Berk Canberk}
[S'03, M'11, SM'16] (B.Canberk@napier.ac.uk; canberk@itu.edu.tr) is a professor within the School of Computing, Engineering and The Build Environment, Edinburgh Napier University, and he is also a professor within the Department of Artificial Intelligence and Data Engineering, Istanbul Technical University. He serves as an Editor in IEEE Transactions on Vehicular Technology, Elsevier Computer Networks and Elsevier Computer Communications. He is the recipient of IEEE Turkey Research Incentive Award (2018), IEEE INFOCOM Best Paper Award (2018), The British Council (UK) Researcher Link Award (2017), IEEE CAMAD Best Paper Award (2016), Royal Society International Exchanges (UK) (2021), Royal Academy of Engineering (UK) NEWTON Research Collaboration Award (2015), IEEE INFOCOM Best Poster Paper Award (2015), ITU Successful Faculty Member Award (2015) and Turkish Telecom Collaborative Research Award (2013). His current research areas are AI-enabled Digital Twins, IoT Communication, and Smart Wireless Networks.
\end{IEEEbiographynophoto}

\end{document}